\documentclass[sigconf,natbib=true]{acmart}

\usepackage{multirow}
\usepackage{multicol}
\usepackage{booktabs}
\usepackage{subcaption}
\usepackage{caption}
\usepackage{setspace}
\usepackage{amsmath}
\usepackage{tabularx}
\usepackage{balance}
\usepackage{textcomp}

\copyrightyear{2022} 
\acmYear{2022} 
\setcopyright{rightsretained} 

\acmConference[SIGIR '22]{Proceedings of the 45th International ACM SIGIR Conference on Research and Development in Information Retrieval}{July 11--15, 2022}{Madrid, Spain.}
\acmBooktitle{Proceedings of the 45th Int'l ACM SIGIR Conference on Research and Development in Information Retrieval (SIGIR '22), July 11--15, 2022, Madrid, Spain}
\acmDOI{10.1145/3477495.3531786}
\acmISBN{978-1-4503-8732-3/22/07}
\usepackage{etoolbox}
\makeatletter
\patchcmd{\maketitle}{\@copyrightpermission}{
   \begin{minipage}{0.3\columnwidth}
     \href{http://creativecommons.org/licenses/by/4.0/}{\includegraphics[width=0.90\textwidth]{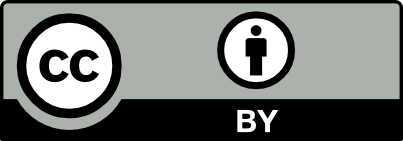}}
   \end{minipage}\hfill
   \begin{minipage}{0.7\columnwidth}
     \href{http://creativecommons.org/licenses/by/4.0/}{This work is licensed under a Creative Commons Attribution International 4.0 License.}
   \end{minipage}
  
   \vspace{5pt}
}{}{}

\makeatother

\definecolor{midnightgreen}{rgb}{0.0, 0.29, 0.33}
\definecolor{darkpink}{rgb}{0.91, 0.33, 0.5}

\newcommand{\ours}{\texttt{P${}^{3}$} \texttt{Ranker}}
\newcommand{\tvanilla}{\text{T5 (Vanilla)}}
\newcommand{\tlm}{\text{T5 (LM-Adapt)}}
\newcommand{\tmulti}{\text{T5 (Multi-task)}}
\newcommand{\mtvanilla}{\text{monoT5 (Vanilla)}}
\begin{document}

\fancyhead{} 

% \title{Understand the Impact of Pretraining-Downstream Gaps on Prompt-Based Learning for Information Retrieval}
\title{P${}^3$ Ranker: Mitigating the Gaps between Pre-training and Ranking Fine-tuning with Prompt-based Learning and Pre-finetuning}

% \author{Xiaomeng Hu$^{1*}$, Shi Yu$^{2,3*}$, Chenyan Xiong$^4$, Zhenghao Liu$^{1\#}$, Zhiyuan Liu$^{2,3\#}$, and Ge Yu$^1$} \thanks{$*$ Equal contribution. \\ $\#$ Corresponding authors.}
% % \affiliation{Northeastern University$^1$, \\ Tsinghua University$^2$, \\ Microsoft Research$^3$ 
% % } 
% \affiliation{$^1$School of Computer Science and Engineering, Northeastern University, Shenyang, China\\ 
% $^2$Department of Computer Science and Technology, Institute for AI, Tsinghua University, Beijing, China\\
% $^3$Beijing National Research Center for Information Science and Technology, China \\
% $^4$Microsoft Research, Redmond, USA
% } 
% \affiliation{
% \texttt{greghxm@foxmail.com};
% \texttt{yus21@mails.tsinghua.edu.cn};
% \texttt{chenyan.xiong@microsoft.com}
% }
% \affiliation{
% \texttt{\{liuzhenghao, yuge\}@mail.neu.edu.cn};
% \texttt{liuzy@tsinghua.edu.cn}
% }

\author{Xiaomeng Hu}
\authornote{Equal contribution.}
\affiliation{%
  \institution{Northeastern University}
  \city{Shenyang}
  \country{China}}
\email{greghxm@foxmail.com}

\author{Shi Yu}
\authornotemark[1]
\affiliation{%
  \institution{Tsinghua University}
  \city{Beijing}
  \country{China}}
\email{yus21@mails.tsinghua.edu.cn}

\author{Chenyan Xiong}
\affiliation{%
  \institution{Microsoft Research}
  \city{Redmond}
  \country{USA}}
\email{chenyan.xiong@microsoft.com}

\author{Zhenghao Liu}
\authornote{Corresponding authors.}
\affiliation{%
  \institution{Northeastern University}
  \city{Shenyang}
  \country{China}}
\email{liuzhenghao@mail.neu.edu.cn}

\author{Zhiyuan Liu}
\authornotemark[2]
\affiliation{%
  \institution{Tsinghua University}
  \city{Beijing}
  \country{China}}
\email{liuzy@tsinghua.edu.cn}

\author{Ge Yu}
\affiliation{%
  \institution{Northeastern University}
  \city{Shenyang}
  \country{China}}
\email{yuge@mail.neu.edu.cn}




\begin{abstract}
Compared to other language tasks, applying pre-trained language models (PLMs) for search ranking often requires more nuances and training signals. 
In this paper, we identify and study the two mismatches between pre-training and ranking fine-tuning: the \textit{training schema gap} regarding the differences in training objectives and model architectures, and the \textit{task knowledge gap} considering the discrepancy between the knowledge needed in ranking and that learned during pre-training.
To mitigate these gaps, we propose Pre-trained, Prompt-learned and Pre-finetuned Neural Ranker (\ours{}). 
\ours{} leverages prompt-based learning to convert the ranking task into a pre-training like schema and uses pre-finetuning to initialize the model on intermediate supervised tasks.
Experiments on MS MARCO and Robust04 show the superior performances of \ours{} in few-shot ranking.
Analyses reveal that \ours{} is able to better accustom to the ranking task through prompt-based learning and retrieve necessary ranking-oriented knowledge gleaned in pre-finetuning, resulting in data-efficient PLM adaptation.
Our code is available at \url{https://github.com/NEUIR/P3Ranker}.

\end{abstract}

\maketitle

\section{Introduction}
\begin{figure}
    \centering
    \includegraphics[width=0.99\linewidth]{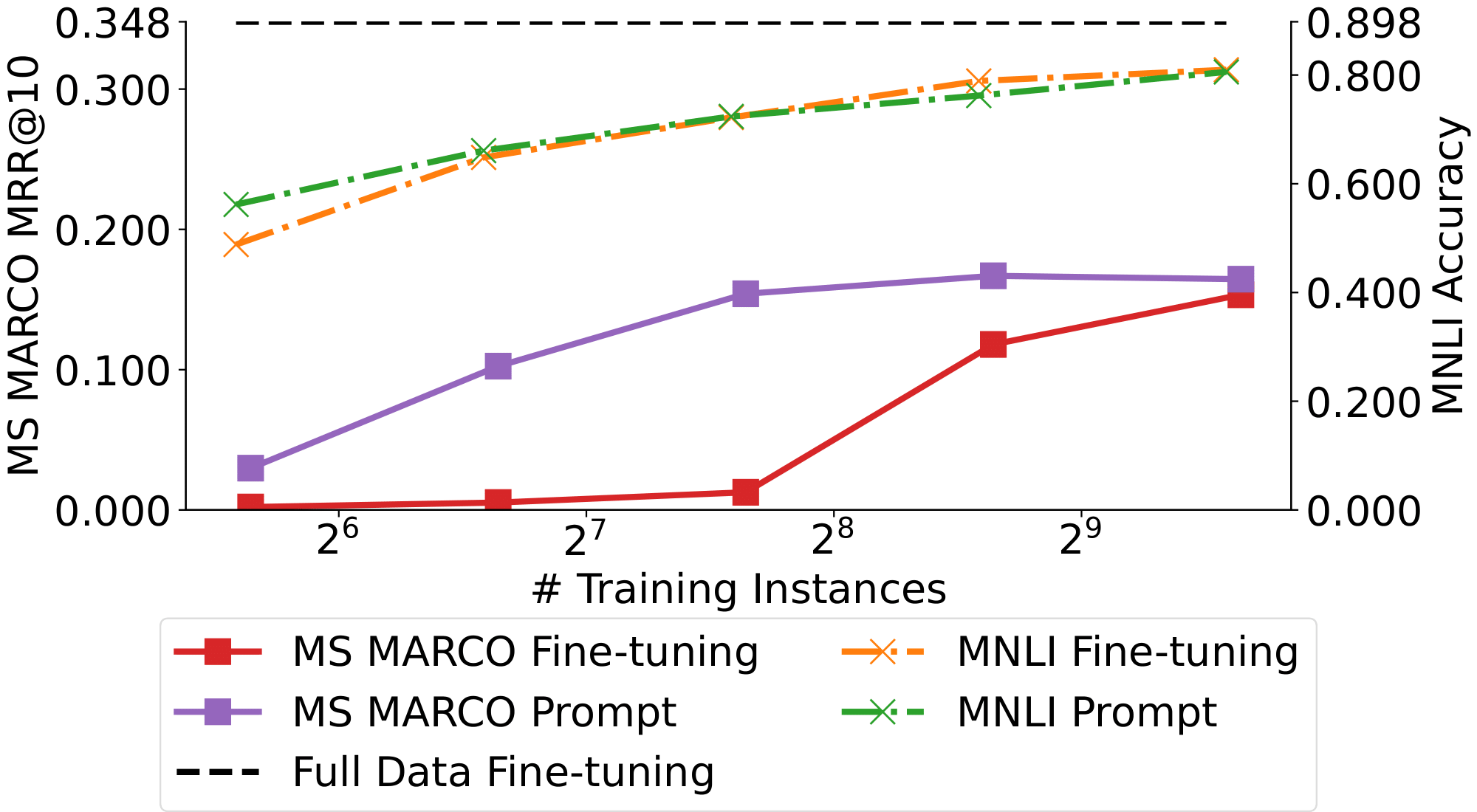}
    \caption{
    Performance of RoBERTa-large on an ad hoc ranking benchmark MS MARCO Passage Ranking and a pairwise NLP benchmark MNLI with vanilla fine-tuning and prompt-based learning.
    % The X-axis denotes the scale of training data.
    % The maximum values in the Y-axes are the performance of the model fine-tuned on full data.
    }
    \label{fig:introduction}
\end{figure}

Recent research applies pre-trained language models (PLMs) for ad hoc search through fine-tuning with plenty of human-annotated relevance pairs~\cite{dai2019deeper, nogueira2019passage, DBLP:journals/corr/abs-1904-07531}.
Attaining outstanding results in data-sufficient scenarios, however, PLMs often showcase poor performance in search ranking scenarios where training labels are limited~\cite{zhang2020selective, DBLP:conf/acl/SunQLXZBLB20}, in contrast to their strong few-shot ability in NLP tasks. 
% As shown in Figure~\ref{fig:introduction}, on an ad hoc ranking dataset MS MARCO~\cite{DBLP:conf/nips/NguyenRSGTMD16}, fine-tuned RoBERTa achieves less than 1\% of the performance of full data fine-tuning when there are $\sim 2^6$ training instances.
% By contrast, on an NLP dataset MultiNLI~\cite{williams2017broad}, with a similar amount of data, the model can achieve over half of the performance of full data fine-tuning.
As shown in Figure~\ref{fig:introduction}, when trained with $\sim 2^6$ instances, RoBERTa achieves less than 1\% of the performance of the full data fine-tuned model on the ranking task MS MARCO~\cite{DBLP:conf/nips/NguyenRSGTMD16} but over half of the performance of the full data fine-tuned model on the NLP task MNLI~\cite{williams2017broad}.
The inability of adapting PLMs into few-shot ranking hinders PLM's application in many real-world search scenarios where relevance labels are expensive to accumulate.

In this paper, we argue that the less optimal few-shot performance of PLM-based ranking models may reside in the unfilled \textit{gaps} between pre-training and search ranking.
Typical PLMs are not tailored for text ranking, leading to two potential gaps.
First, there is a \textit{training schema gap}: pre-training typically uses token-level objectives like masked language modeling (MLM), but ranking fine-tuning uses a sequence-pair level classification objective and needs to learn a new MLP head from scratch.
Second, there also exists a \textit{task knowledge gap}: The knowledge and signals gleaned from unsupervised LM pre-training may not benefit the ranking task, which needs to extract the underlying relevance signals on the $(q, d)$ pairs~\cite{ma2021prop, boualili2020markedbert, xiong2017end}. 
The two gaps, one from the schema perspective and another from the knowledge perspective, challenge PLMs' adaptation into ranking tasks in few-shot scenarios.

This work mitigates the gaps with a Pre-trained, Prompt-learned and Pre-finetuned Neural Ranker (\ours{}). 
\ours{} uses the \textit{prompt-based learning} technique, casting the input into a ``prompt'' and obtaining output from token predictions~\cite{DBLP:conf/nips/BrownMRSKDNSSAA20,lester2021power, DBLP:conf/acl/GaoFC20, DBLP:journals/corr/abs-2103-10385}.
This ensures a consistent input/output form and maintains the same model architecture as pre-training during ranking fine-tuning, closing the training schema gap.
Using prompt-based learning, as depicted in Figure~\ref{fig:introduction}, boost the performance on MS MARCO, but still can't match the few-shot advantage on MNLI.
% This indicates that using prompt-based learning is not enough to fully address the mismatch.
% ; the second gap is also critical for search ranking.
We argue that the remaining discrepancy results from the lack of ranking knowledge in PLMs, i.e. the existence of the task knowledge gap.
% We infer that the remaining 
To mitigate the gap, we introduce \textit{pre-finetuning} to warm up the model on intermediate NLP tasks before final fine-tuning.
The pre-finetuning stage serves as a ``bridge'' between unsupervised pre-training and ranking fine-tuning where the model can learn useful knowledge for ranking. 

Experiments on ad hoc ranking benchmarks MS MARCO~\cite{DBLP:conf/nips/NguyenRSGTMD16} and Robust04~\cite{kwok2004trec} show that \ours{} has strong advantages in few-shot ranking. 
On MS MARCO, when trained on only 5 queries, \ours{} retains 45\% of the performance of full-data fine-tuning, greatly outperforming vanilla pre-training/fine-tuning baselines.
Analyses find that both prompt-based learning and pre-finetuning contribute to \ours{}'s effectiveness.
By closely resembling the pre-training form, prompt-based learning methods better unleash the power of PLMs for search ranking, narrowing the training schema gap.
Pre-finetuning, on the other hand, offers ranking-oriented knowledge which can be easily transformed and leveraged in few-shot scenarios, narrowing the task knowledge gap.

\section{Related Work}
This section reviews previous studies on PLM-based search ranking, prompt-based learning and pre-finetuning.

\textbf{PLM-based Search Ranking.} 
Search ranking models have been developed and thrived on plenty of NLP tasks, e.g., question answering~\cite{chen2017reading,lee2019latent,DBLP:conf/sigir/Qu0CQCI20} and fact verification~\cite{DBLP:conf/acl/LiuXSL20}. 
Recent models are usually based on pre-trained language models (PLMs)~\cite{DBLP:journals/corr/abs-1904-07531,dai2019deeper,DBLP:conf/emnlp/NogueiraJPL20,nogueira2019passage,DBLP:journals/corr/abs-1910-14424}.
In a typical reranking setting, every candidate $(q, d)$ pair from a first-stage retrieval system is concatenated and fed into a PLM, which outputs a relevance score for ranking.
Though PLMs have significantly improved ranking performance with large amounts of data, they tend to be less competitive with traditional lexical models in few-shot ranking scenarios~\cite{DBLP:conf/acl/SunQLXZBLB20,zhang2020selective}.
% However, existing language model pretraining methods are unfortunate to broaden their advantages to IR scenarios, making deep Neu-IR models even need more training signals than shallow ones. 

\textbf{Prompt-based Learning.}
Recently, prompt-based learning~\cite{DBLP:conf/nips/BrownMRSKDNSSAA20,DBLP:journals/corr/abs-2107-13586} is proposed to adapt language model to downstream tasks.
Unlike traditional fine-tuning, prompt-based learning reformulates model input to make downstream task closer to pre-training through a \textit{template}.  
% At the core of prompt-based learning is the choice of \textit{templates}.
A template consists of a series of prompt words for the model to understand the task, typically defined manually~\cite{DBLP:conf/eacl/SchickS21} or generated automatically~\cite{DBLP:conf/emnlp/ShinRLWS20,DBLP:conf/acl/GaoFC20}.
The model learns to fill label words in the reserved blank in the template as output.
Thanks to prompt-based learning, PLMs can easily adapt to downstream tasks and achieve admirable performance with few labeled data on a wide range of NLP tasks~\cite{DBLP:conf/eacl/SchickS21,DBLP:conf/acl/GaoFC20,DBLP:conf/emnlp/ShinRLWS20}. 
Nevertheless, it is still unclear whether it can benefit PLM-based ranking models and alleviate the gaps between language model pre-training and ranking fine-tuning.

\textbf{Pre-finetuning.} 
Researchers have explored the idea of pre-finetuning on single/multiple supervised intermediate task(s) in order to improve the downstream performance of PLMs~\cite{vu2020exploring, aghajanyan2021muppet, pruksachatkun2020intermediate, wang2019can, talmor2019multiqa}. 
Pre-finetuned models generally show better performance on downstream tasks and are more data-efficient~\cite{vu2020exploring, aghajanyan2021muppet}. 
However, the criteria for selecting intermediate tasks remain unclear, and unsuitable ones can bring negative impacts~\cite{pruksachatkun2020intermediate}. 
Inspired by previous research, our work explores the effectiveness of pre-finetuning on search ranking together with prompt-based learning.
% introduce IR and recent methods
% introduce prompt and point out its usage in IR

\section{Methodology}
\label{sec:method}

The ad hoc ranking task is to estimate a relevance score $P(y|(q,d))$ for a user query $q$ and a document $d$ from a document collection. 

Recent ranking models follow the pre-training/fine-tuning paradigm, encoding the query and document together and adding a layer on the top to perform classification.
Those models rely highly on the in-domain relevance labels for training, limiting the performance in few-shot scenarios.
To mitigate the transfer gaps, we propose \ours{} to start from a pre-trained model and then use the method of prompt-based learning (Section~\ref{subsec:prompt}) to perform pre-finetuning (Section~\ref{subsec:prefinetuning}) and final fine-tuning.
% is optimized with prompt-based learning to closely resemble pre-training (Section~\ref{subsec:prompt}).
% Pre-finetuning is introduced for \ours{} to learn ranking-related knowledge for later transfer (Section~\ref{subsec:prefinetuning}).

% \subsection{Vanilla Fine-tuning for IR}
% Previous work~\cite{dai2019deeper,DBLP:journals/corr/abs-1904-07531} uses encoder-only transformer-based model like BERT for ranking. The input is the concatenation of a query $q$ and document $d$ from a training sample $X=(q,d)$:
% \begin{equation}
% \small
%     \mathbf{H}=\text{BERT}(\text{[CLS]}~~q~~\text{[SEP]}~~d~~\text{[SEP]}),
% \end{equation}
% We take the representation of the [CLS] token, $\mathbf{h}_{\text{[CLS]}}$ from $\mathbf{H}$ and use it to calculate the relevance of a query-document pair $<q,d>$:
% \begin{equation}
% \small
%     P(y|<q,d>) = \text{softmax}_y (\mathbf{W}\mathbf{h}_{\text{[CLS]}}),
% \end{equation}
% where the relevance label $y\in \mathcal{Y}$ represents either relevant ($y=1$) or irrelevant ($y=0$).

\subsection{Prompt-based Learning for Search Ranking}
\ours{} alleviates the training schema gap between pre-training and fine-tuning by using prompt-based learning methods.
This is done with a predefined template $T$ and a verbalizer $\mathcal{M}$. 

\textbf{Building Input with a Template.} 
\label{subsec:prompt}
A template consists of slots for input data and several prompt tokens serving as hints to the PLM.
\ours{} is based on the encoder-decoder model T5~\cite{DBLP:journals/corr/abs-1910-10683}.
We follow ~\citet{DBLP:conf/emnlp/NogueiraJPL20} to reformulate a $(q,d)$ pair with the following template:
\begin{equation}
\small
    T(q,d)=\text{Query:}~~\texttt{[q]}~~\text{Document:}~~\texttt{[d]}~~\text{Relevant:}~~,
\end{equation}
where \texttt{[q]}, \texttt{[d]} are slots for the query and document.
The $(q,d)$ pair is filled in and then fed into the model.
% After filling the slots with corresponding texts, the T5 encoder is used to encode the inputs.

\textbf{Obtaining Output from Label Words.} 
\ours{} sniffs the output directly from output label words.
To achieve this, a verbalizer $\mathcal{M}(y)$ is used to establish the mapping from the task label space $\mathcal{Y}$ to the label word space $\mathcal{V}$.
The verbalizer of \ours{} maps relevant ($y=1$) to ``true'' and irrelevant ($y=0$) to ``false''.
The final prediction is based on the softmax over the set of label words:
\begin{align}
\small
    P(y|(q,d))&=P(t=\mathcal{M}(y))=\frac{\exp (\mathbf{w}_{\mathcal{M}(y)}^{\mathsf{T}} \mathbf{h}_{t})}{\sum_{y^{\prime} \in \mathcal{Y}} \exp (\mathbf{w}_{\mathcal{M}(y^{\prime})}^{\mathsf{T}} \mathbf{h}_{t})},
\end{align}
where $t$, $\mathbf{h}_t$ denote the first token from the decoder and its hidden representation. $\mathbf{w}$ is the corresponding vector in the language modeling head. 
During training, we use the cross entropy (CE) loss.

\subsection{Pre-finetuning for Search Ranking}
\label{subsec:prefinetuning}
To mitigate the task knowledge gap, we introduce an additional \textit{pre-finetuning} process in \ours{} between LM pre-training and fine-tuning.
By pre-finetuning, we expect \ours{} to inherit the knowledge from the massive labels of other supervised NLP tasks, serving as a complement to the scarce labels in few-shot ranking scenarios.
Starting from a PLM, we continuously train the model on language understanding tasks using prompt-based learning with manually defined task-specific templates and verbalizers.
Then the pre-trained and pre-finetuned model is ready for final prompt-based fine-tuning on ranking datasets.
% We select pairwise language understanding tasks for pre-finetuning.
% We manually define task-specific templates and verbalizers for pre-finetuning tasks, and continuously train T5 on those tasks keeping a consistent loss as in Section~\ref{subsec:prompt}.
% We choose MultiNLI (MNLI) as the task for \ours{} in our main experiments for pre-finetuning. 
% The MNLI task is to predict the entailment information between two sentences.
% With the help of prompt-based mothods, the input is manipulated into:
% \begin{equation}
%     \text{mnli hypothesis:}~~\texttt{[h]}~~\text{precise:}~~\texttt{[p]}~~\text{entailment:},
% \end{equation}
% and the verbalizer maps ``entailment'', ``neutral'' and ``contradiction'' labels to ``true'', ``neutral'' and ``false'' respectively.
% Our pre-finetuning is based on the prompt-based learning paradigm, 

\section{Experimental Methodology}
\label{sec:exp_method}
This section describes our experimental settings and implementation details.

\textbf{Datasets and Metrics.} 
We train and evaluate our models on MS MARCO Passage~\cite{DBLP:conf/nips/NguyenRSGTMD16} and Robust04~\cite{kwok2004trec}.
MS MARCO is a large-scale IR dataset containing 530k training queries. 
The evaluation of MS MARCO is on the official dev set and the evaluation metric is MRR@10.
Robust04 contains TREC-style fine-grained annotations on 249 queries and we use NDCG@20 as the evaluation metric.
% We do five-fold cross-validation with the evaluation metric NDCG@20.

\textbf{Few-Shot Configurations.}
To simulate different training scenarios, we partition the training data according to training queries on MS MARCO and relevance labels on Robust04:
On MS MARCO, for there just exists 1 relevant passage per query, we sample \{5, 50, 1k, 530k (all)\} training queries from the training set, each paired with a relevant passage and an irrelevant passage returned by BM25;
On Robust04, for there are far more than 1 relevance label (often >1k) per query, we directly sample \{0.2\%, 2\%, 100\%\} relevance labels.

\begin{table*}[t]
    \centering
    \small
    \caption{Overall results on MS MARCO official Dev set and Robust04 under different training data sizes. We sample training data based on the numbers of queries on MS MARCO and relevance labels on Robust04. See Section~\ref{sec:exp_method} for more details. $\dagger$, $\ddagger$, $\mathsection$, and $\mathparagraph$ indicate statistically significant improvements over Lexical$\text{}^{\dagger}$, BERT$\text{}^{\ddagger}$, RoBERTa$\text{}^{\mathsection}$, and monoT5$\text{}^{\mathparagraph}$, respectively.}
    % \resizebox{\linewidth}{!}{
     \begin{tabular}{l|l|l|l|l|l|l|l} 
    % \hline
    \hline
    \multirow{2}{*}{\textbf{Model}} & \multicolumn{4}{c|}{\textbf{MS MARCO MRR@10}} & \multicolumn{3}{c}{\textbf{Robust04 NDCG@20}} \\
    \cline{2-8}
                                    & \textbf{5 Queries} & \textbf{50 Queries}    & \textbf{1k Queries}    & \textbf{All Queries (530k)}    & \textbf{0.2\% Labels} & \textbf{2\% Labels}   &  \textbf{All Labels (100\%)} \\
    \hline
    \textbf{Lexical}                         & \multicolumn{4}{c|}{\textbf{0.1874} (BM25)}                                                                       & \multicolumn{3}{c}{\textbf{0.4269} (SDM)}   \\
    \hline
    \multicolumn{8}{l}{\textbf{Base Models}} \\
    \hline
    BERT                            & 0.0038$\text{}^{\mathsection \mathparagraph}$
                                    & 0.1035$\text{}^{\ddagger \mathsection}$
                                    & \textbf{0.2210}$\text{}^{\dagger \mathsection \mathparagraph}$
                                    & 0.3518$\text{}^\dagger \mathparagraph$

                                    & 0.2563$\text{}^{\mathsection \mathparagraph}$                
                                    & 0.3178$\text{}^{\mathsection \mathparagraph}$
                                    & 0.4625$\text{}^{\dagger \mathparagraph}$\\

    RoBERTa                         & 0.0011            
                                    & 0.0064$\text{}^{\mathparagraph}$                
                                    & 0.1771$\text{}^{\mathparagraph}$                
                                    & 0.3471$\text{}^{\dagger \mathparagraph}$                       

                                    & 0.2054                
                                    & 0.2171                
                                    & \textbf{0.4634}$\text{}^{\dagger \mathparagraph}$\\

    monoT5~\cite{DBLP:conf/emnlp/NogueiraJPL20}                     & 0.0009            
                                    & 0.0037
                                    & 0.0157                
                                    & 0.3351$\text{}^{\dagger}$

                                    & 0.1645                
                                    & 0.1669                
                                    & 0.2269\\
    PROP~\cite{ma2021prop}          & \textbf{0.1198}$\text{}^{\ddagger \mathsection \mathparagraph}$
                                    & \textbf{0.1656}$\text{}^{\ddagger \mathsection \mathparagraph}$
                                    & 0.2098$\text{}^{\dagger \mathsection \mathparagraph}$
                                    & \textbf{0.3519}$\text{}^{\dagger \mathparagraph}$
                                    
                                    & \textbf{0.3484}$\text{}^{\ddagger \mathsection \mathparagraph}$
                                    & \textbf{0.3967}$\text{}^{\ddagger \mathsection \mathparagraph}$
                                    & 0.4569$\text{}^{\dagger \mathparagraph}$ \\

    \ours{}                         & 0.0523$\text{}^{\ddagger \mathsection \mathparagraph}$
                                    & 0.0949$\text{}^{\mathsection \mathparagraph}$
                                    & 0.2027$\text{}^{\dagger \mathsection \mathparagraph}$                
                                    & 0.3311$\text{}^{\dagger}$

                                    & 0.2410$\text{}^{\mathsection \mathparagraph}$                
                                    & 0.3292$\text{}^{\mathsection \mathparagraph}$
                                    & 0.4403$\text{}^{\mathparagraph}$ \\
    \hline
    \multicolumn{8}{l}{\textbf{Large Models}} \\
    \hline
    BERT                            & 0.0245$\text{}^{\mathsection \mathparagraph}$            
                                    & 0.1296$\text{}^{\mathsection \mathparagraph}$
                                    & 0.2259$\text{}^{\dagger \mathparagraph}$                
                                    & 0.3526$\text{}^{\dagger}$ 
                                    
                                    & 0.3196$\text{}^{\mathsection \mathparagraph}$                
                                    & 0.3801$\text{}^{\mathsection \mathparagraph}$                
                                    & 0.4888$\text{}^{\dagger}$\\
                                    
    RoBERTa                         & 0.0017            
                                    & 0.0053                
                                    & 0.2570$\text{}^{\dagger \ddagger \mathparagraph}$                
                                    & 0.3475$\text{}^{\dagger}$                         
                                    
                                    & 0.2089$\text{}^{\mathparagraph}$                
                                    & 0.1933$\text{}^{\mathparagraph}$               
                                    & 0.5086$\text{}^{\dagger \ddagger}$ \\
                                    
    monoT5~\cite{DBLP:conf/emnlp/NogueiraJPL20}& 0.0029$\text{}^{\mathsection}$ 
                                    & 0.1117$\text{}^{\mathsection}$
                                    & 0.1616                
                                    & 0.3496$\text{}^{\dagger}$
                                    
                                    & 0.1386                
                                    & 0.1424                
                                    & 0.2271\\
                                    
    \ours{}                         & \textbf{0.1659}$\text{}^{\ddagger \mathsection \mathparagraph}$            
                                    & \textbf{0.1943}$\text{}^{\dagger \ddagger \mathsection \mathparagraph}$                 
                                    & \textbf{0.2876}$\text{}^{\dagger \ddagger \mathsection \mathparagraph}$    
                                    & \textbf{0.3645}$\text{}^{\dagger \ddagger \mathsection \mathparagraph}$               
                                    
                                    & \textbf{0.3787}$\text{}^{\ddagger \mathsection \mathparagraph}$                
                                    & \textbf{0.4321}$\text{}^{\ddagger \mathsection \mathparagraph}$             
                                    & \textbf{0.5213}$\text{}^{\dagger \ddagger \mathsection \mathparagraph}$ \\
                                    
    % \ours{}-large (Anchor)                &                   
    %                                 & 0.0516                
    %                                 &                       
    %                                 &                               
                                    
    %                                 &                       
    %                                 &                       
    %                                 & \\
    \hline
    % \hline
    \end{tabular} 
    \label{tab:overall-results-msmarco}
\end{table*}

% \textbf{Models.} 
% We compare BERT, RoBERTa and T5 for ranking~\cite{DBLP:journals/corr/abs-1904-07531,dai2019deeper,DBLP:conf/emnlp/NogueiraJPL20}. 
% For BERT and RoBERTa, we include vanilla fine-tuning, prompt-based learning with manually and automatically designed prompts, as well as continuous prompts.
% For Seq2Seq model T5, we use manually designed prompts and 

% It further comes up with a simple prompt template and estimate query-document relevance by predicting label word probabilities. Two variants of T5 %with and without multitask training are considered as baselines.
% that with different pre-training objectives are considered as baselines, too.

\textbf{Baselines.} 
We compare \ours{} to vanilla fine-tuning models BERT and RoBERTa as well as monoT5~\cite{DBLP:conf/emnlp/NogueiraJPL20} in our main experiments.
MonoT5 is a T5-based ranker trained with manually defined prompts, but it does not have the pre-finetuning process.
To have a fair comparison, monoT5 is initialized with the vanilla pre-trained version (denoted as \tvanilla{}), i.e. the one that is pre-trained solely on the unsupervised denoising objective~\cite{DBLP:journals/corr/abs-1910-10683}. 
We also compare our model with PROP~\cite{ma2021prop}, a ranking-oriented pre-training model.
For lexical retrieval models, we report Anserini BM25~\cite{yang2017anserini} on MS MARCO and SDM ~\cite{metzler2005markov} on Robust04.

\textbf{Implementation Details.} 
In our main experiments, we initialize \ours{} with \tvanilla{} and pre-finetune it on MNLI~\cite{williams2017broad} for 12k steps with batch size set to 32.
The batch size of the fine-tuning process is set to 8 when there are less than 50 queries on MS MARCO and 2\% labels on Robust04, and 32 for remaining cases.
During inference, we rerank top 1000 passages from Anserini BM25 on MS MARCO and top 100 passages from SDM on Robust04.
We do five-fold cross-validation on Robust04.
In both the pre-finetuning and fine-tuning stages we use Adam as the optimizer with the initial learning rate set to 2e-5 and linear decay.

\section{Evaluation Results}

\begin{table}
    \centering
    \small
    \caption{Comparison of prompt-based methods and vanilla fine-tuning. All experiments are conducted using the large models on MS MARCO. The evaluation metric is MRR@10.}
    \resizebox{\linewidth}{!}{
     \begin{tabular}{l|l|l|l|l|l} 
    \hline
    \multirow{2}{*}{\textbf{Model}} & \multirow{2}{*}{\textbf{Training Schema}}   & \multicolumn{4}{c}{\textbf{\# Training Queries}}  \\
    \cline{3-6}
                                    &                                           & \textbf{5q}   & \textbf{50q}  & \textbf{1k q} & \textbf{All q}\\ 
    \hline
    \multirow{3}{*}{BERT}           & Vanilla Fine-tune                         & \textbf{0.0245}& \textbf{0.1296}& \textbf{0.2259}        & \textbf{0.3526} \\
                                    & Manual Prompt                             & 0.0038        & 0.0385        & 0.2181        & 0.3413 \\
                                    & Auto Prompt                               & 0.0043        & 0.0289        & 0.2218        & 0.3471  \\
    \hline
    \multirow{3}{*}{RoBERTa}        & Vanilla Fine-tune                         & 0.0017        & 0.0053        & 0.2570        & \textbf{0.3475} \\
                                    & Manual Prompt                             & 0.0043        & 0.1025        & 0.2547        & 0.3439 \\
                                    & Auto Prompt                               & \textbf{0.0062}        & \textbf{0.1806}        & \textbf{0.2789}        & 0.3412 \\
    \hline
    \multirow{2}{*}{monoT5}             & No-word Prompt                            & 0.0020             & 0.0111             & 0.0775             & 0.3470       \\
                                    & Manual Prompt                             & \textbf{0.0029}        & \textbf{0.1117}        & \textbf{0.1616}        & \textbf{0.3496}  \\
    \hline
    \end{tabular} }
    % \caption{Comparison of prompt design methods. All experiments use RoBERTa-Large.}
    \label{tab:prompt-methods}

\end{table}
In this section, we present overall results and the analyses of prompt-based learning and pre-finetuning.

\subsection{Overall Results}

The results on MS MARCO and Robust04 are presented in Table~\ref{tab:overall-results-msmarco}.

\ours{} demonstrates strong few-shot ability on MS MARCO: It can acquire basic search ranking ability with only 5 training queries, outperforming all PLM-based ranking models except PROP by an order of magnitude.
Note that our approach has a different intention to PROP: We explore pre-finetuning the model with supervised NLP labels, whereas PROP employs ranking-oriented pre-training with plenty of generated $(q,d)$ pairs.
The large version of \ours{} has a significant performance boost, outperforming all neural baselines among all data sizes.
On Robust04, the large version of \ours{} also outperforms neural baselines by large margins in few-shot scenarios.

A usable neural ranker should be able to improve over traditional lexical retrieval models. 
To achieve this goal, \ours{} only needs to use 50 queries on MS MARCO or 2\% relevance labels on Robust04, whereas baseline models require much more data.
That makes \ours{} applicable to special ranking domains where relevance labels are difficult to accumulate.

\subsection{Effectiveness of Prompt-based Learning}
To have a thorough understanding of the effectiveness of prompt-based learning on mitigating the training schema gap, in this experiment, we compare BERT, RoBERTa, and T5 with different prompt schemes as well as vanilla fine-tuning\footnote{For encoder-only models BERT and RoBERTa, the template include a [MASK] token for model prediction. See Appendix~\ref{appendix:prompt} for more details.}. 
The prompt schemes that we consider include manually designed and automatically designed prompts.
For manually designed prompts (``Manual Prompt''), the template is manually defined according to human understanding of the task.
For automatically designed prompts (``Auto Prompt''), we borrow the methods from LM-BFF~\cite{DBLP:conf/acl/GaoFC20}, which utilizes T5 to automatically generate prompts. 
Results are presented in Table~\ref{tab:prompt-methods}.

Reformulating the task into a prompt format helps minimize the training schema gap for RoBERTa but not for BERT.
The contrasting results may stem from the different pre-training schemes -- RoBERTa is just pre-trained with MLM whereas BERT has an extra pre-training scheme, Next Sentence Prediction (NSP).
In addition, RoBERTa is pre-trained with MLM for much longer steps than BERT, having the potential to better understand the clues in the prompt.
Compared to manually designed prompts, automatically designed prompts seem to be more effective for RoBERTa.

Different from BERT and RoBERTa, T5 processes pre-training and all downstream tasks in a unified text-to-text framework with manually designed prompts, thus there is no training schema gap during its' downstream adaption.
We compare the performance of manual prompt with a ``No-word Prompt'' version, where all prompt words are removed from the input.
As shown in Table~\ref{tab:prompt-methods}, the performance downgrades greatly after removing prompt words.
Maintaining a proper form of prompts is beneficial for the model to understand the task, leading to better performance.

%--------------------------------------

\subsection{Effectiveness of Pre-finetuning}

In this experiment, we study the effectiveness of pre-finetuning on mitigating the task knowledge gap by investigating the performance of pre-finetuning on different tasks and steps. 
To better showcase how well the ranking-oriented knowledge is learned during pre-finetuning, we present model performance with ``prompt tuning''~\cite{lester2021power}, where we fix all the parameters of the PLM and only tune a few continuous token embeddings, which account for less than 1\% of total parameters.
In this way, adapting to the downstream task introduces little additional knowledge.
Thus better prompt tuning performance indicates more ranking-oriented knowledge learned during pre-finetuning.
% In this way, adapting to the downstream task introduces limited additional knowledge and thus the model is pushed to exploit what it has already learned during pre-finetuning.
% This method is often referred to as ``prompt tuning''~\cite{lester2021power}.
To distinguish, the method that we use normally to train \ours{} is referred to as ``model tuning''.

\begin{figure}[t]
    \centering
    \begin{subfigure}[t]{0.49\columnwidth}
        \centering
        \includegraphics[width=\linewidth]{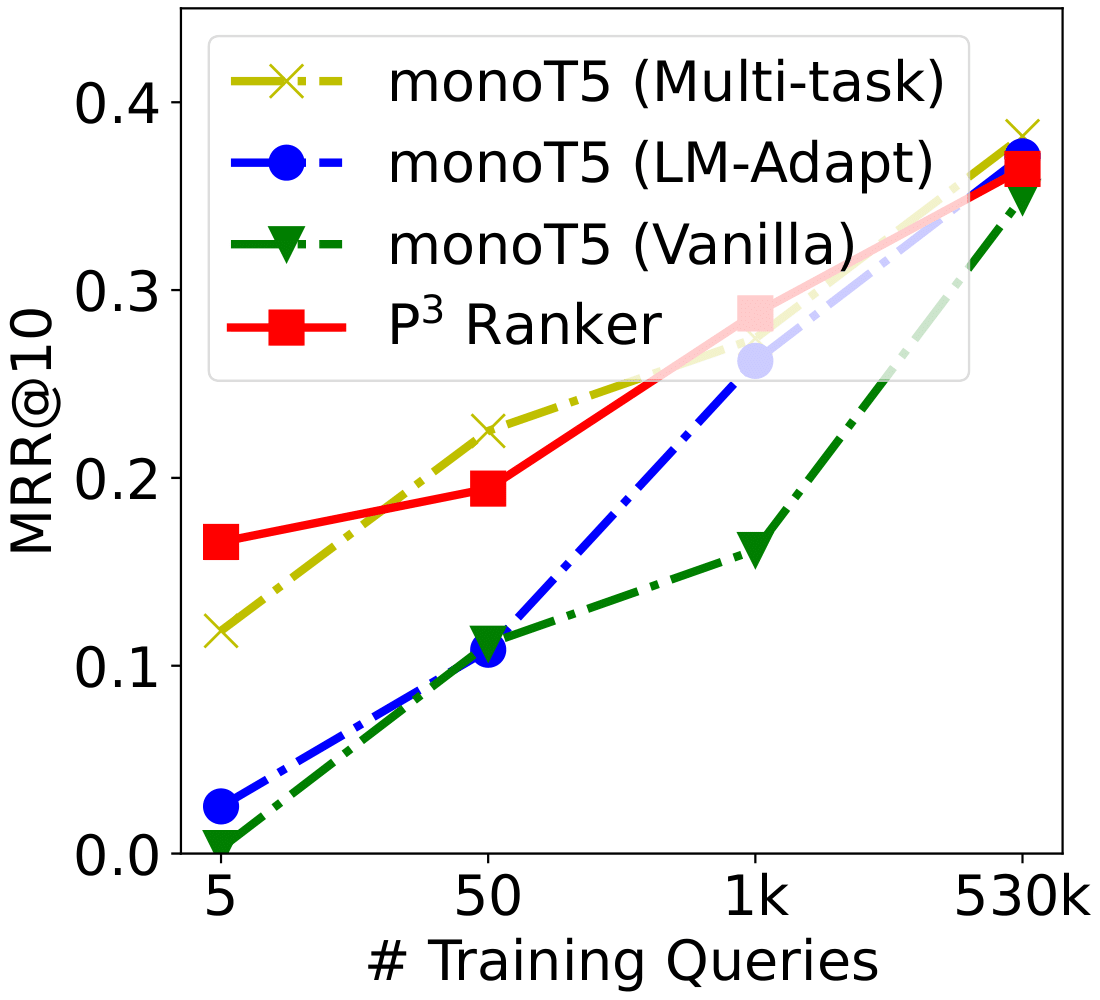}
        \caption{Model Tuning. \label{fig:t5-ablation-finetune}}
    \end{subfigure}
    \begin{subfigure}[t]{0.49\columnwidth}
        \centering
        \includegraphics[width=\linewidth]{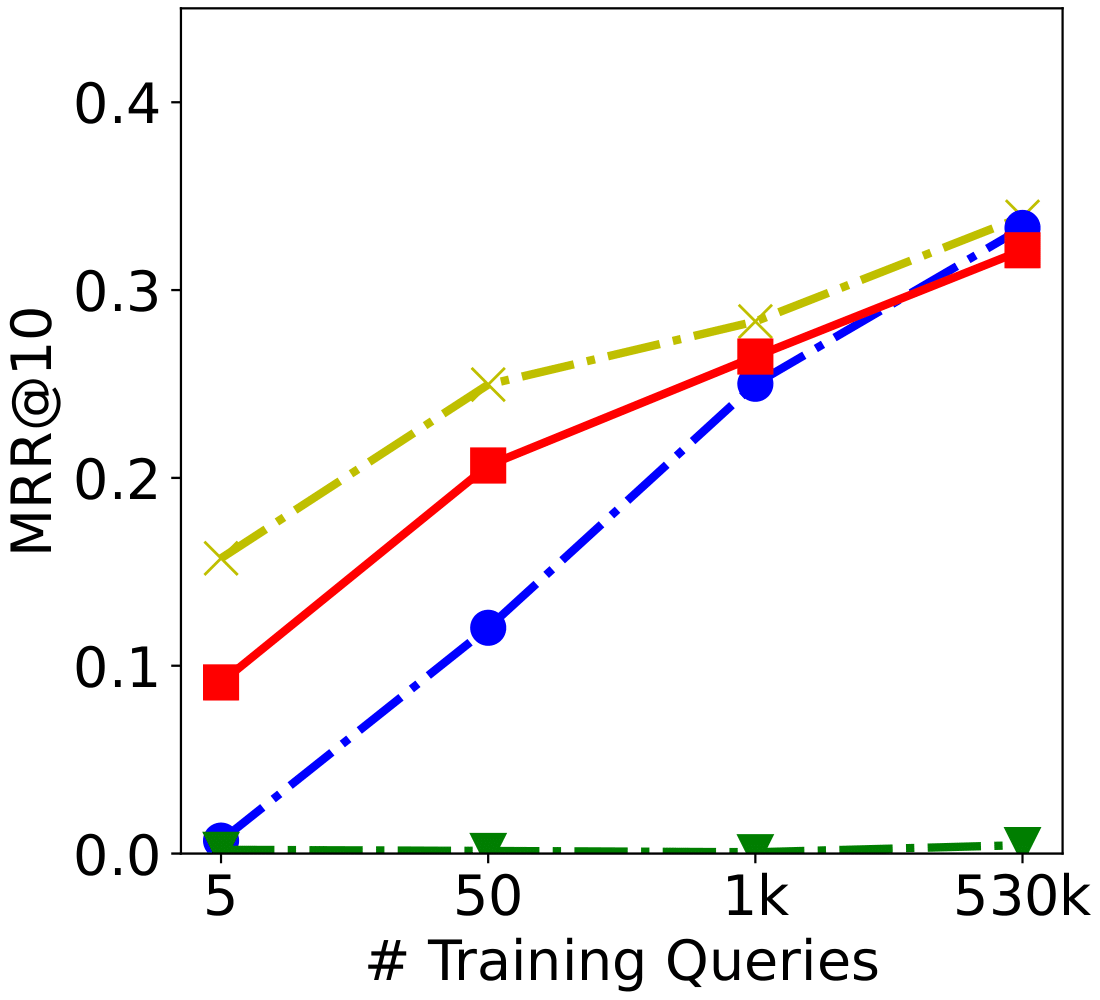}
       
        \caption{Prompt Tuning. \label{fig:t5-ablation-softprompt}}
    \end{subfigure}
    \caption{Comparisons of the ranking performance of \ours{} and different monoT5 models. Experiments are conducted on MS MARCO.
    \label{fig:t5-ablation}}
\end{figure}

\textbf{Performance Gains from Pre-finetuning.}
As shown in Figure~\ref{fig:t5-ablation}, we compare the performance of \ours{} to monoT5 initialized from three officially released pre-trained T5 checkpoints: \tvanilla{}, \tlm{} and \tmulti{}.
\tlm{} is pre-finetuned using the prefix LM objective~\cite{lester2021power} and \tmulti{} is pre-trained in a multi-task manner~\cite{DBLP:journals/corr/abs-1910-10683}.
As shown in Figure~\ref{fig:t5-ablation-finetune}, \ours{} shows on-par performance with mono\tmulti{}, and it greatly outperforms other models in few-shot scenarios.
Mono\tmulti{} benefits from knowledge of multiple task sources, but with an overhead of extensive pre-training on various tasks with careful mixing techniques.
Compared to mono\tmulti{}, \ours{} is simpler and more efficient for it only needs pre-finetuning on one task for only several thousands of steps.
In the case of prompt tuning, as shown in Figure~\ref{fig:t5-ablation-softprompt}, \ours{} still yields results close to mono\tmulti{}.
By pre-finetuning on a single task, though keeping the parameters of the PLM fixed, \ours{} is still able to rank documents through exploiting the knowledge obtained during pre-finetuning, which mono\tlm{} and mono\tvanilla{} are unable to leverage in few-shot scenarios.
% The poor performance of mono\tlm{} and mono\tvanilla{} indicates that vanilla LM pre-training objectives are unable to introduce ranking related knowledge.

\begin{figure}[t]
    \centering
    \begin{subfigure}[t]{0.49\columnwidth}
        \centering
        \includegraphics[width=\linewidth]{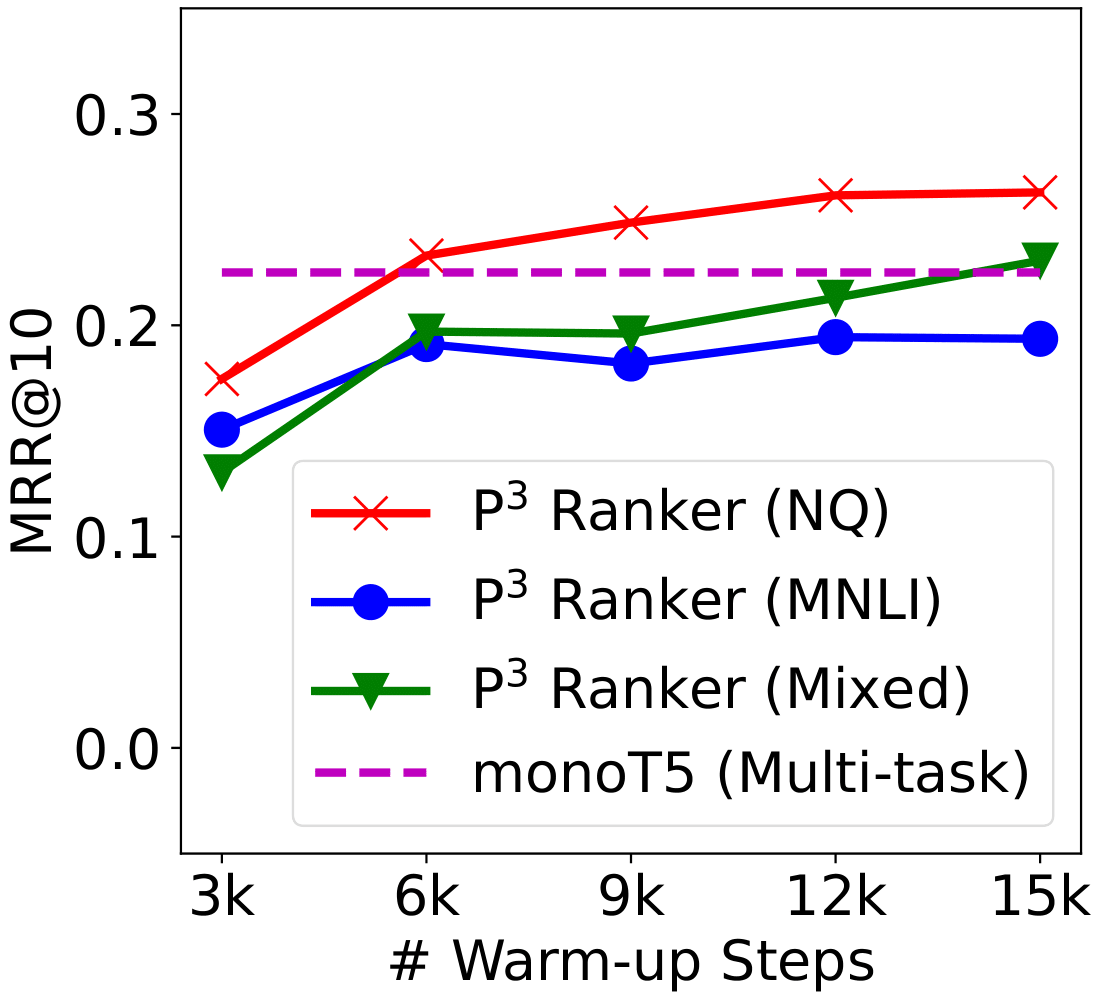}
        \caption{Model Tuning. \label{fig:warm_up_finetune}}
    \end{subfigure}
    \begin{subfigure}[t]{0.49\columnwidth}
        \centering
        \includegraphics[width=\linewidth]{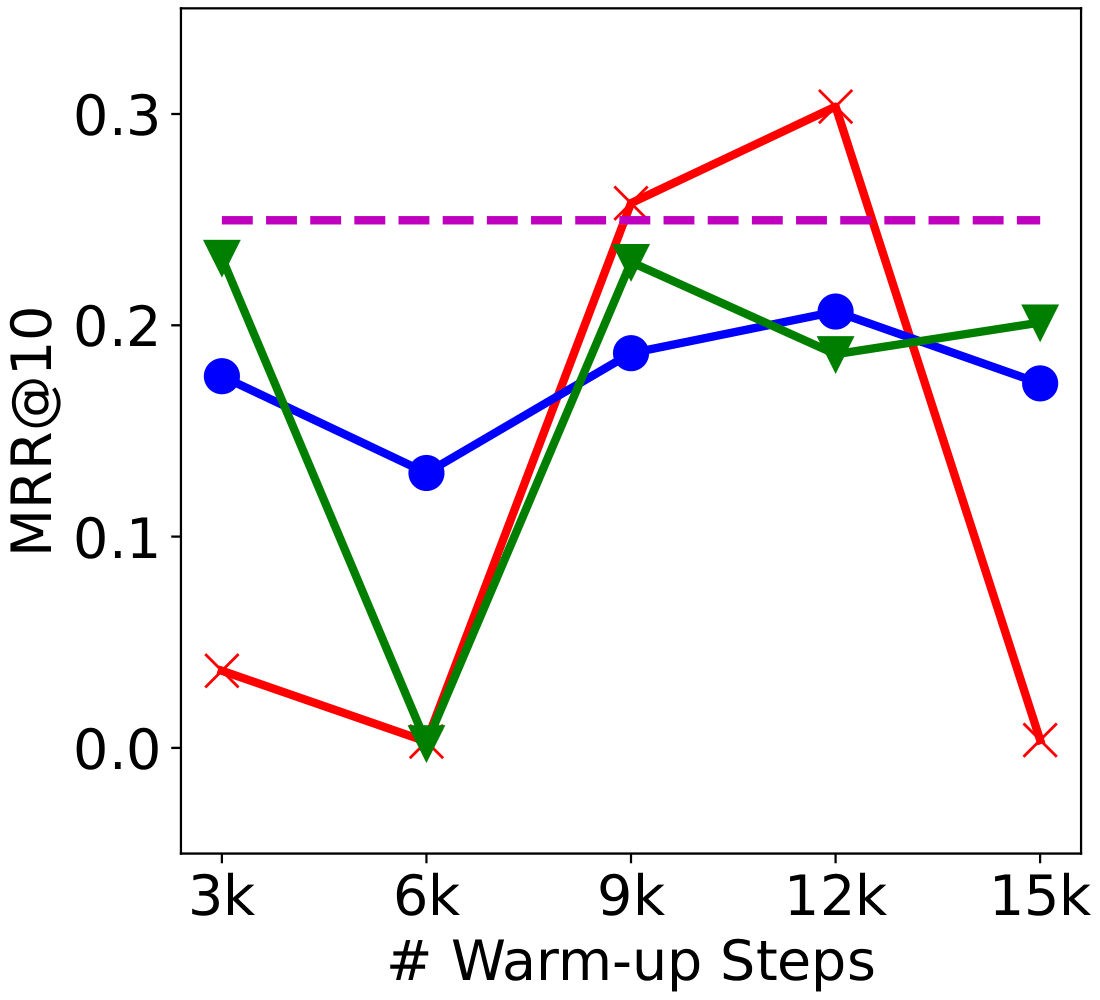}
        \caption{Prompt Tuning. \label{fig:warm_up_softprompt}}
    \end{subfigure}
    \caption{Ranking performance on MS MARCO after pre-finetuning on intermediate tasks. All models are fine-tuned with only 50 queries.
    \label{fig:Warm-up with tasks}}
\end{figure}

\textbf{Task and Model Selection for Pre-finetuning.}
As shown in Figure~\ref{fig:Warm-up with tasks}, we compare the performance of \ours{} pre-finetuned with other tasks and under varying steps.
We introduce a pre-finetuning task on Natural Questions (NQ) to predict the answerability of a passage given a question.
We also conduct multi-task learning on an equal mixture of the MNLI and NQ task. 
We find that using a single task, either MNLI or NQ, can effectively narrow the knowledge gap.
The NQ-pre-finetuned version tends to yield better performance; 
The answerability-checking task NQ seems to be more similar to ranking, a relevance checking task, than the entailment prediction task MNLI.
The mixture of the two tasks doesn't bring significant improvements.
Multi-task pre-finetuning for ranking needs further study.

%\begin{figure}[h]
  %  \centering
    %\includegraphics[width=\linewidth]{Figures/task-pretrain.pdf}
 %   \includegraphics[width=\linewidth]{Figures/bar_pretrain.pdf}
%    \caption{Task Pretrain.
%    \label{fig:Task Pretrain}}
%\end{figure}
% \textbf{Embedding Analysis.}
\subsection{Knowledge Transfer}
\begin{figure}
    \centering
    \begin{subfigure}[t]{0.45\columnwidth}
        \centering
        \includegraphics[width=\linewidth]{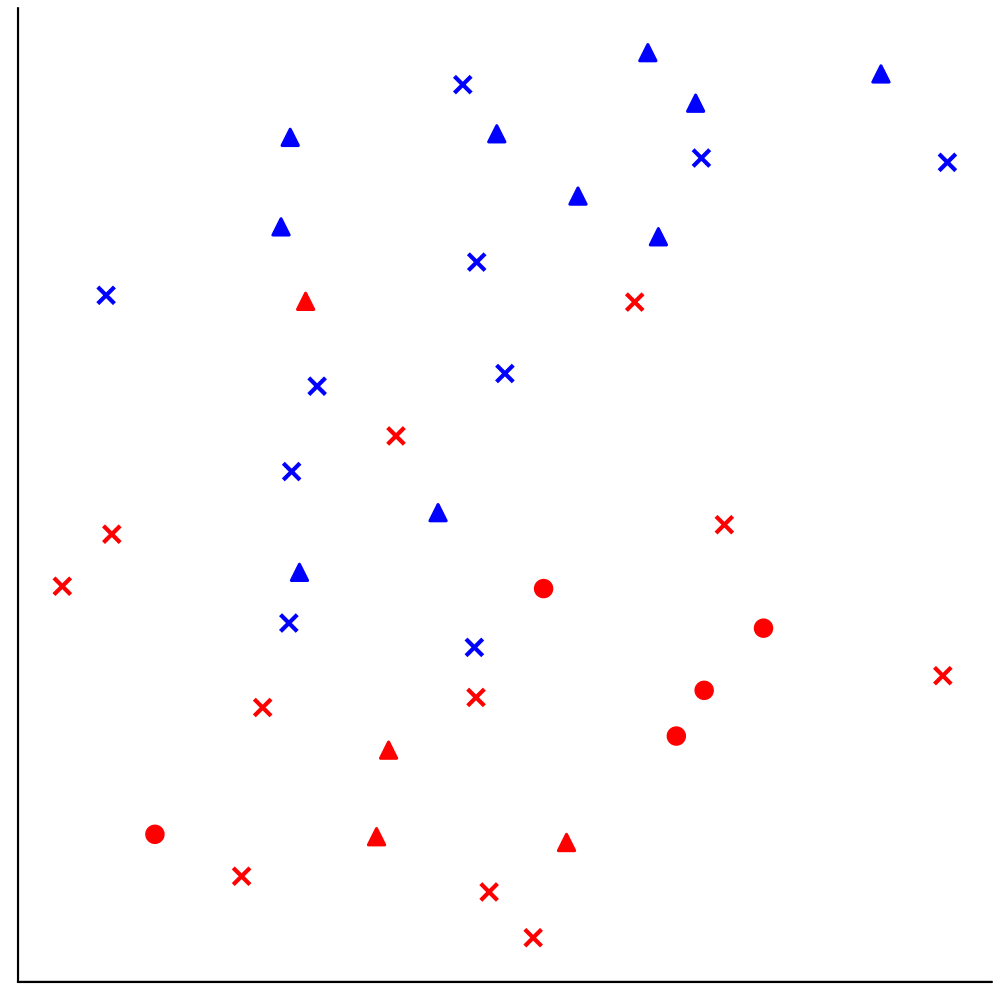}
        \caption{\mtvanilla{}. \label{fig:t5-v11-tsne}}
    \end{subfigure}
    \begin{subfigure}[t]{0.45\columnwidth}
        \centering
        \includegraphics[width=\linewidth]{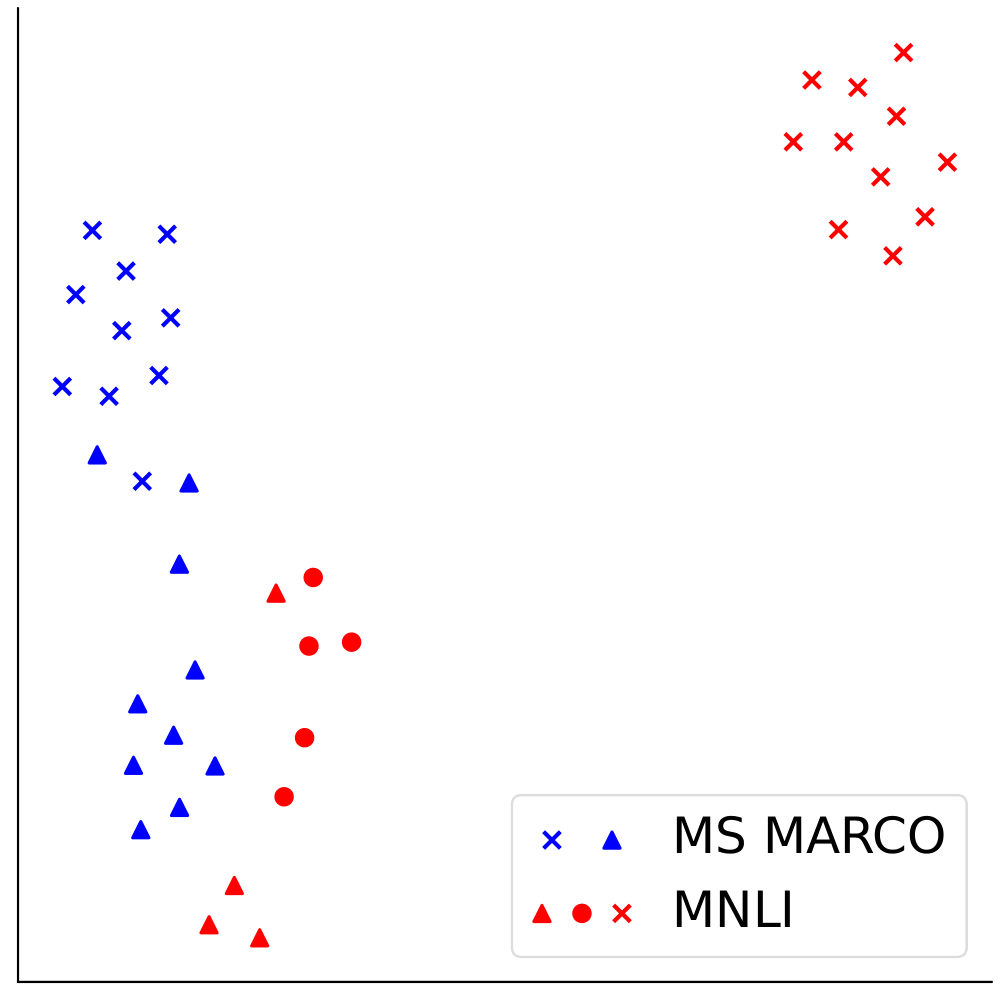}
        \caption{\ours{}. \label{fig:t5-v11-warmup-tsne}}
    \end{subfigure}
    \caption{T-SNE visualizes the embeddings of the token used for classification from (a) \mtvanilla{} and (b) \ours{}. Both models are trained with prompt tuning on 50 training queries. Blue triangles and crosses denote MS MARCO positive pairs and negative pairs, respectively. Red triangles, dots, and crosses denote MNLI ``entailment'', ``neutral'' and ``contradiction'' pairs, respectively.
    \label{fig:t5-tsne}}
\end{figure}
\begin{figure}
    \centering
    \begin{subfigure}[t]{0.49\columnwidth}
        \centering
        \includegraphics[width=\linewidth]{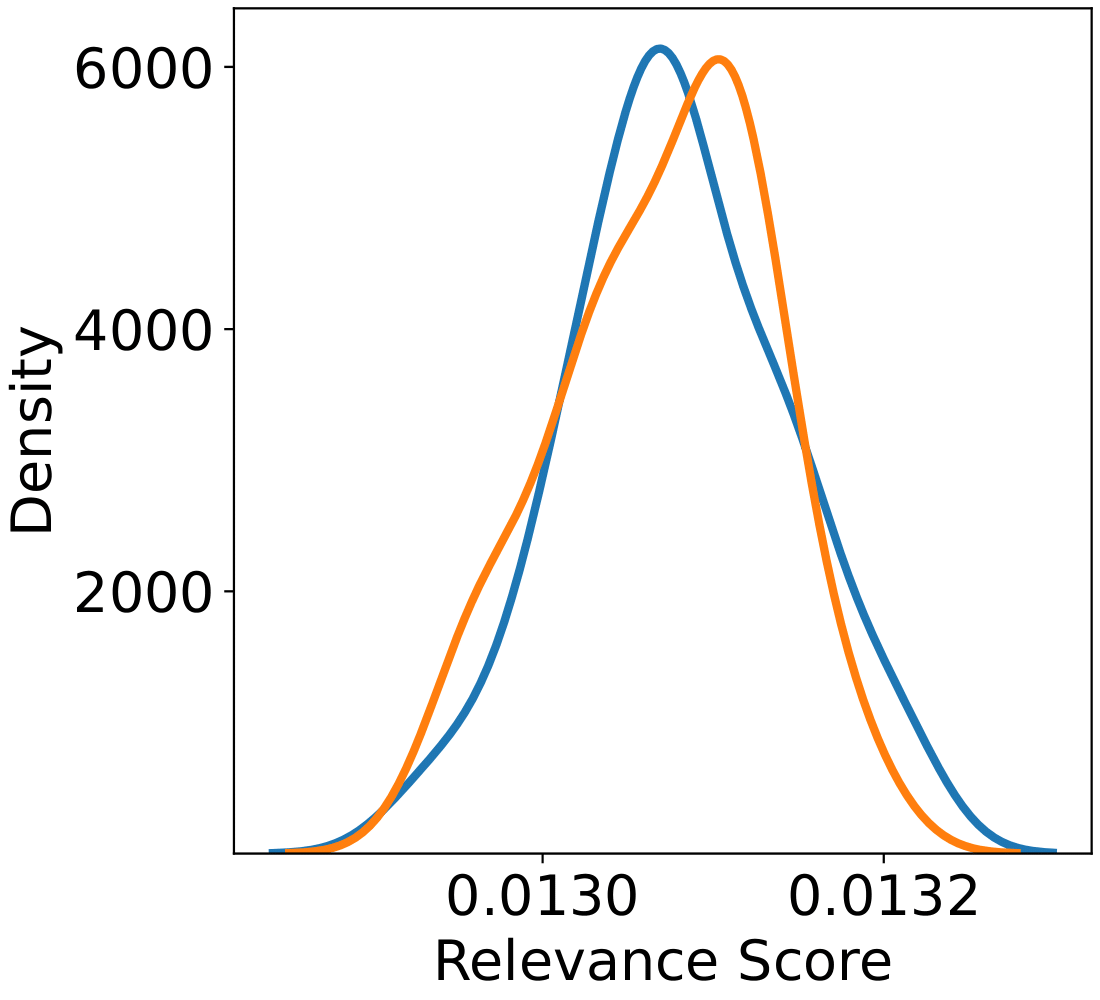}
        \caption{\mtvanilla{}. \label{fig:t5-v11-dist}}
    \end{subfigure}
    \begin{subfigure}[t]{0.49\columnwidth}
        \centering
        \includegraphics[width=\linewidth]{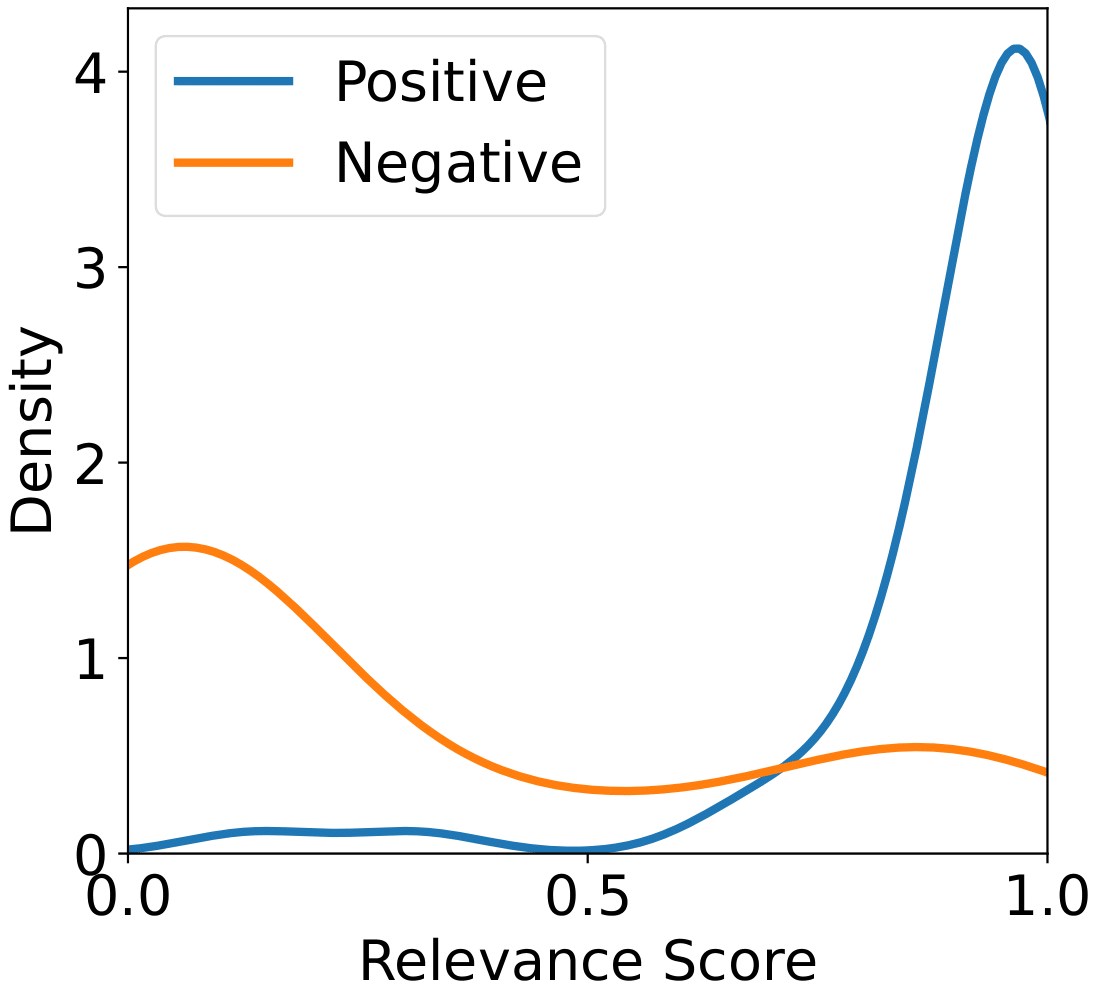}
        \caption{\ours{}. \label{fig:t5-v11-warmup-dist}}
    \end{subfigure}
    \caption{Relevance score distributions of positive and negative (q,d) pairs on (a) \mtvanilla{} and (b) \ours{}. Both models are trained with prompt tuning on 50 queries.
    \label{fig:t5-score}}
\end{figure}
In this experiment, we study the embeddings of the token used for prediction to figure out how the knowledge gleaned in pre-finetuning is transferred into search ranking. 
In Figure~\ref{fig:t5-tsne}, we plot the t-SNE visualizations of the embeddings produced by \ours{} and \mtvanilla{}.
Without pre-finetuning, the embeddings from mono\tvanilla{} distribute randomly, and the positive/negative $(q,d)$ pairs cannot be distinguished.
After pre-finetuning, the positive/negative pairs are separated, and the positive pairs seem to take advantage of the knowledge from MNLI neutral/entailment pairs.
The distributions of the relevance scores produced by the two models are presented in Figure~\ref{fig:t5-score}.
Mono\tvanilla{} yields similar scores for the pos/neg pairs, whilst \ours{} learns to assign scores with a large gap with the help of pre-finetuning knowledge.

\section{Conclusion}
In this paper, we present a few-shot ranking model for search ranking, \ours{}. 
\ours{} narrows the training schema gap and task knowledge gap between pre-training and ranking fine-tuning by using prompt-based learning and pre-finetuning.
Our experiments show that \ours{} can effectively learn to rank documents in few-shot scenarios.
We present analyses on the prompt-based learning and pre-finetuning, and find that both of them contribute to mitigating the gaps, leading to a strong overall performance.

\section*{Acknowledgments}
This work is mainly supported by Beijing Academy of Artificial Intelligence (BAAI) as well as supported in part by the Fundamental Research Funds for the Central Universities under Grant N2216013, National Science and Technology Major Project (J2019-IV-0002-0069), the Natural Science Foundation of China under Grant 62006129 and the College Scientific Research Project of Inner Mongolia under Grant NJZY21263.

\balance

% \bibliography{anthology}
% \bibliographystyle{acl_natbib}

\bibliographystyle{ACM-Reference-Format}
\bibliography{anthology}


\begin{thebibliography}{33}

%%% the Web address.

\ifx \showCODEN    \undefined \def \showCODEN     #1{\unskip}     \fi
\ifx \showDOI      \undefined \def \showDOI       #1{#1}\fi
\ifx \showISBNx    \undefined \def \showISBNx     #1{\unskip}     \fi
\ifx \showISBNxiii \undefined \def \showISBNxiii  #1{\unskip}     \fi
\ifx \showISSN     \undefined \def \showISSN      #1{\unskip}     \fi
\ifx \showLCCN     \undefined \def \showLCCN      #1{\unskip}     \fi
\ifx \shownote     \undefined \def \shownote      #1{#1}          \fi
\ifx \showarticletitle \undefined \def \showarticletitle #1{#1}   \fi
\ifx \showURL      \undefined \def \showURL       {\relax}        \fi
% invisible to TeX
\providecommand\bibfield[2]{#2}
\providecommand\bibinfo[2]{#2}
\providecommand\natexlab[1]{#1}
\providecommand\showeprint[2][]{arXiv:#2}

\bibitem[Aghajanyan et~al\mbox{.}(2021)]%
        {aghajanyan2021muppet}
\bibfield{author}{\bibinfo{person}{Armen Aghajanyan}, \bibinfo{person}{Anchit
  Gupta}, \bibinfo{person}{Akshat Shrivastava}, \bibinfo{person}{Xilun Chen},
  \bibinfo{person}{Luke Zettlemoyer}, {and} \bibinfo{person}{Sonal Gupta}.}
  \bibinfo{year}{2021}\natexlab{}.
\newblock \showarticletitle{Muppet: Massive Multi-task Representations with
  Pre-Finetuning}. In \bibinfo{booktitle}{\emph{Proceedings of EMNLP}}.
  \bibinfo{pages}{5799--5811}.
\newblock


\bibitem[Bajaj et~al\mbox{.}(2016)]%
        {DBLP:conf/nips/NguyenRSGTMD16}
\bibfield{author}{\bibinfo{person}{Payal Bajaj}, \bibinfo{person}{Daniel
  Campos}, \bibinfo{person}{Nick Craswell}, \bibinfo{person}{Li Deng},
  \bibinfo{person}{Jianfeng Gao}, \bibinfo{person}{Xiaodong Liu},
  \bibinfo{person}{Rangan Majumder}, \bibinfo{person}{Andrew McNamara},
  \bibinfo{person}{Bhaskar Mitra}, \bibinfo{person}{Tri Nguyen},
  {et~al\mbox{.}}} \bibinfo{year}{2016}\natexlab{}.
\newblock \showarticletitle{MS MARCO: A human generated MAchine Reading
  COmprehension dataset}.
\newblock \bibinfo{journal}{\emph{arXiv preprint arXiv:1611.09268}}
  (\bibinfo{year}{2016}).
\newblock


\bibitem[Boualili et~al\mbox{.}(2020)]%
        {boualili2020markedbert}
\bibfield{author}{\bibinfo{person}{Lila Boualili}, \bibinfo{person}{Jose~G.
  Moreno}, {and} \bibinfo{person}{Mohand Boughanem}.}
  \bibinfo{year}{2020}\natexlab{}.
\newblock \showarticletitle{MarkedBERT: Integrating Traditional {IR} Cues in
  Pre-trained Language Models for Passage Retrieval}. In
  \bibinfo{booktitle}{\emph{Proceedings of SIGIR}}.
  \bibinfo{pages}{1977--1980}.
\newblock


\bibitem[Brown et~al\mbox{.}(2020)]%
        {DBLP:conf/nips/BrownMRSKDNSSAA20}
\bibfield{author}{\bibinfo{person}{Tom~B. Brown}, \bibinfo{person}{Benjamin
  Mann}, \bibinfo{person}{Nick Ryder}, \bibinfo{person}{Melanie Subbiah},
  \bibinfo{person}{Jared Kaplan}, \bibinfo{person}{Prafulla Dhariwal},
  \bibinfo{person}{Arvind Neelakantan}, \bibinfo{person}{Pranav Shyam},
  \bibinfo{person}{Girish Sastry}, \bibinfo{person}{Amanda Askell},
  \bibinfo{person}{Sandhini Agarwal}, \bibinfo{person}{Ariel Herbert{-}Voss},
  \bibinfo{person}{Gretchen Krueger}, \bibinfo{person}{Tom Henighan},
  \bibinfo{person}{Rewon Child}, \bibinfo{person}{Aditya Ramesh},
  \bibinfo{person}{Daniel~M. Ziegler}, \bibinfo{person}{Jeffrey Wu},
  \bibinfo{person}{Clemens Winter}, \bibinfo{person}{Christopher Hesse},
  \bibinfo{person}{Mark Chen}, \bibinfo{person}{Eric Sigler},
  \bibinfo{person}{Mateusz Litwin}, \bibinfo{person}{Scott Gray},
  \bibinfo{person}{Benjamin Chess}, \bibinfo{person}{Jack Clark},
  \bibinfo{person}{Christopher Berner}, \bibinfo{person}{Sam McCandlish},
  \bibinfo{person}{Alec Radford}, \bibinfo{person}{Ilya Sutskever}, {and}
  \bibinfo{person}{Dario Amodei}.} \bibinfo{year}{2020}\natexlab{}.
\newblock \showarticletitle{Language Models are Few-Shot Learners}. In
  \bibinfo{booktitle}{\emph{Processing of NeurIPS}}.
\newblock


\bibitem[Chen et~al\mbox{.}(2017)]%
        {chen2017reading}
\bibfield{author}{\bibinfo{person}{Danqi Chen}, \bibinfo{person}{Adam Fisch},
  \bibinfo{person}{Jason Weston}, {and} \bibinfo{person}{Antoine Bordes}.}
  \bibinfo{year}{2017}\natexlab{}.
\newblock \showarticletitle{Reading Wikipedia to Answer Open-Domain Questions}.
  In \bibinfo{booktitle}{\emph{Proceedings of ACL}}.
  \bibinfo{pages}{1870--1879}.
\newblock


\bibitem[Dai and Callan(2019)]%
        {dai2019deeper}
\bibfield{author}{\bibinfo{person}{Zhuyun Dai} {and} \bibinfo{person}{Jamie
  Callan}.} \bibinfo{year}{2019}\natexlab{}.
\newblock \showarticletitle{Deeper Text Understanding for {IR} with Contextual
  Neural Language Modeling}. In \bibinfo{booktitle}{\emph{Proceedings of
  SIGIR}}. \bibinfo{pages}{985--988}.
\newblock


\bibitem[Gao et~al\mbox{.}(2021)]%
        {DBLP:conf/acl/GaoFC20}
\bibfield{author}{\bibinfo{person}{Tianyu Gao}, \bibinfo{person}{Adam Fisch},
  {and} \bibinfo{person}{Danqi Chen}.} \bibinfo{year}{2021}\natexlab{}.
\newblock \showarticletitle{Making Pre-trained Language Models Better Few-shot
  Learners}. In \bibinfo{booktitle}{\emph{Proceedings of ACL}}.
  \bibinfo{pages}{3816--3830}.
\newblock


\bibitem[Kwok et~al\mbox{.}(2004)]%
        {kwok2004trec}
\bibfield{author}{\bibinfo{person}{Kui-Lam Kwok}, \bibinfo{person}{Laszlo
  Grunfeld}, \bibinfo{person}{HL Sun}, \bibinfo{person}{Peter Deng}, {and}
  \bibinfo{person}{N Dinstl}.} \bibinfo{year}{2004}\natexlab{}.
\newblock \showarticletitle{TREC 2004 Robust Track Experiments Using PIRCS}. In
  \bibinfo{booktitle}{\emph{Text REtrieval Conference (TREC)}}.
\newblock


\bibitem[Lee et~al\mbox{.}(2019)]%
        {lee2019latent}
\bibfield{author}{\bibinfo{person}{Kenton Lee}, \bibinfo{person}{Ming-Wei
  Chang}, {and} \bibinfo{person}{Kristina Toutanova}.}
  \bibinfo{year}{2019}\natexlab{}.
\newblock \showarticletitle{Latent Retrieval for Weakly Supervised Open Domain
  Question Answering}. In \bibinfo{booktitle}{\emph{Proceedings of ACL}}.
  \bibinfo{pages}{6086--6096}.
\newblock


\bibitem[Lester et~al\mbox{.}(2021)]%
        {lester2021power}
\bibfield{author}{\bibinfo{person}{Brian Lester}, \bibinfo{person}{Rami
  Al-Rfou}, {and} \bibinfo{person}{Noah Constant}.}
  \bibinfo{year}{2021}\natexlab{}.
\newblock \showarticletitle{The Power of Scale for Parameter-Efficient Prompt
  Tuning}. In \bibinfo{booktitle}{\emph{Proceedings of EMNLP}}.
  \bibinfo{pages}{3045--3059}.
\newblock


\bibitem[Liu et~al\mbox{.}(2021a)]%
        {DBLP:journals/corr/abs-2107-13586}
\bibfield{author}{\bibinfo{person}{Pengfei Liu}, \bibinfo{person}{Weizhe Yuan},
  \bibinfo{person}{Jinlan Fu}, \bibinfo{person}{Zhengbao Jiang},
  \bibinfo{person}{Hiroaki Hayashi}, {and} \bibinfo{person}{Graham Neubig}.}
  \bibinfo{year}{2021}\natexlab{a}.
\newblock \showarticletitle{Pre-train, prompt, and predict: A systematic survey
  of prompting methods in natural language processing}.
\newblock \bibinfo{journal}{\emph{arXiv preprint arXiv:2107.13586}}
  (\bibinfo{year}{2021}).
\newblock


\bibitem[Liu et~al\mbox{.}(2021b)]%
        {DBLP:journals/corr/abs-2103-10385}
\bibfield{author}{\bibinfo{person}{Xiao Liu}, \bibinfo{person}{Yanan Zheng},
  \bibinfo{person}{Zhengxiao Du}, \bibinfo{person}{Ming Ding},
  \bibinfo{person}{Yujie Qian}, \bibinfo{person}{Zhilin Yang}, {and}
  \bibinfo{person}{Jie Tang}.} \bibinfo{year}{2021}\natexlab{b}.
\newblock \showarticletitle{GPT understands, too}.
\newblock \bibinfo{journal}{\emph{arXiv preprint arXiv:2103.10385}}
  (\bibinfo{year}{2021}).
\newblock


\bibitem[Liu et~al\mbox{.}(2020)]%
        {DBLP:conf/acl/LiuXSL20}
\bibfield{author}{\bibinfo{person}{Zhenghao Liu}, \bibinfo{person}{Chenyan
  Xiong}, \bibinfo{person}{Maosong Sun}, {and} \bibinfo{person}{Zhiyuan Liu}.}
  \bibinfo{year}{2020}\natexlab{}.
\newblock \showarticletitle{Fine-grained Fact Verification with Kernel Graph
  Attention Network}. In \bibinfo{booktitle}{\emph{Proceedings of ACL}}.
  \bibinfo{pages}{7342--7351}.
\newblock


\bibitem[Ma et~al\mbox{.}(2021)]%
        {ma2021prop}
\bibfield{author}{\bibinfo{person}{Xinyu Ma}, \bibinfo{person}{Jiafeng Guo},
  \bibinfo{person}{Ruqing Zhang}, \bibinfo{person}{Yixing Fan},
  \bibinfo{person}{Xiang Ji}, {and} \bibinfo{person}{Xueqi Cheng}.}
  \bibinfo{year}{2021}\natexlab{}.
\newblock \showarticletitle{PROP: Pre-training with Representative Words
  Prediction for Ad-hoc Retrieval}. In \bibinfo{booktitle}{\emph{Proceedings of
  WSDM}}. \bibinfo{pages}{283--291}.
\newblock


\bibitem[Metzler and Croft(2005)]%
        {metzler2005markov}
\bibfield{author}{\bibinfo{person}{Donald Metzler} {and}
  \bibinfo{person}{W~Bruce Croft}.} \bibinfo{year}{2005}\natexlab{}.
\newblock \showarticletitle{A Markov random field model for term dependencies}.
  In \bibinfo{booktitle}{\emph{Proceedings of SIGIR}}.
  \bibinfo{pages}{472--479}.
\newblock


\bibitem[Nogueira and Cho(2019)]%
        {nogueira2019passage}
\bibfield{author}{\bibinfo{person}{Rodrigo Nogueira} {and}
  \bibinfo{person}{Kyunghyun Cho}.} \bibinfo{year}{2019}\natexlab{}.
\newblock \showarticletitle{Passage Re-ranking with BERT}.
\newblock \bibinfo{journal}{\emph{arXiv preprint arXiv:1901.04085}}
  (\bibinfo{year}{2019}).
\newblock


\bibitem[Nogueira et~al\mbox{.}(2020)]%
        {DBLP:conf/emnlp/NogueiraJPL20}
\bibfield{author}{\bibinfo{person}{Rodrigo Nogueira}, \bibinfo{person}{Zhiying
  Jiang}, \bibinfo{person}{Ronak Pradeep}, {and} \bibinfo{person}{Jimmy Lin}.}
  \bibinfo{year}{2020}\natexlab{}.
\newblock \showarticletitle{Document Ranking with a Pretrained
  Sequence-to-Sequence Model}. In \bibinfo{booktitle}{\emph{Findings of the
  ACL: EMNLP 2020}}. \bibinfo{pages}{708--718}.
\newblock


\bibitem[Nogueira et~al\mbox{.}(2019)]%
        {DBLP:journals/corr/abs-1910-14424}
\bibfield{author}{\bibinfo{person}{Rodrigo Nogueira}, \bibinfo{person}{Wei
  Yang}, \bibinfo{person}{Kyunghyun Cho}, {and} \bibinfo{person}{Jimmy Lin}.}
  \bibinfo{year}{2019}\natexlab{}.
\newblock \showarticletitle{Multi-stage document ranking with BERT}.
\newblock \bibinfo{journal}{\emph{arXiv preprint arXiv:1910.14424}}
  (\bibinfo{year}{2019}).
\newblock


\bibitem[Pruksachatkun et~al\mbox{.}(2020)]%
        {pruksachatkun2020intermediate}
\bibfield{author}{\bibinfo{person}{Yada Pruksachatkun}, \bibinfo{person}{Jason
  Phang}, \bibinfo{person}{Haokun Liu}, \bibinfo{person}{Phu~Mon Htut},
  \bibinfo{person}{Xiaoyi Zhang}, \bibinfo{person}{Richard~Yuanzhe Pang},
  \bibinfo{person}{Clara Vania}, \bibinfo{person}{Katharina Kann}, {and}
  \bibinfo{person}{Samuel Bowman}.} \bibinfo{year}{2020}\natexlab{}.
\newblock \showarticletitle{Intermediate-Task Transfer Learning with Pretrained
  Language Models: When and Why Does It Work?}. In
  \bibinfo{booktitle}{\emph{Proceedings of ACL}}. \bibinfo{pages}{5231--5247}.
\newblock


\bibitem[Qiao et~al\mbox{.}(2019)]%
        {DBLP:journals/corr/abs-1904-07531}
\bibfield{author}{\bibinfo{person}{Yifan Qiao}, \bibinfo{person}{Chenyan
  Xiong}, \bibinfo{person}{Zhenghao Liu}, {and} \bibinfo{person}{Zhiyuan Liu}.}
  \bibinfo{year}{2019}\natexlab{}.
\newblock \showarticletitle{Understanding the Behaviors of BERT in Ranking}.
\newblock \bibinfo{journal}{\emph{arXiv preprint arXiv:1904.07531}}
  (\bibinfo{year}{2019}).
\newblock


\bibitem[Qu et~al\mbox{.}(2020)]%
        {DBLP:conf/sigir/Qu0CQCI20}
\bibfield{author}{\bibinfo{person}{Chen Qu}, \bibinfo{person}{Liu Yang},
  \bibinfo{person}{Cen Chen}, \bibinfo{person}{Minghui Qiu},
  \bibinfo{person}{W~Bruce Croft}, {and} \bibinfo{person}{Mohit Iyyer}.}
  \bibinfo{year}{2020}\natexlab{}.
\newblock \showarticletitle{Open-retrieval conversational question answering}.
  In \bibinfo{booktitle}{\emph{Proceedings of SIGIR}}.
  \bibinfo{pages}{539--548}.
\newblock


\bibitem[Raffel et~al\mbox{.}(2020)]%
        {DBLP:journals/corr/abs-1910-10683}
\bibfield{author}{\bibinfo{person}{Colin Raffel}, \bibinfo{person}{Noam
  Shazeer}, \bibinfo{person}{Adam Roberts}, \bibinfo{person}{Katherine Lee},
  \bibinfo{person}{Sharan Narang}, \bibinfo{person}{Michael Matena},
  \bibinfo{person}{Yanqi Zhou}, \bibinfo{person}{Wei Li}, {and}
  \bibinfo{person}{Peter~J. Liu}.} \bibinfo{year}{2020}\natexlab{}.
\newblock \showarticletitle{Exploring the Limits of Transfer Learning with a
  Unified Text-to-Text Transformer}.
\newblock \bibinfo{journal}{\emph{J. Mach. Learn. Res.}}  \bibinfo{volume}{21}
  (\bibinfo{year}{2020}), \bibinfo{pages}{140:1--140:67}.
\newblock


\bibitem[Schick and Sch{\"u}tze(2021)]%
        {DBLP:conf/eacl/SchickS21}
\bibfield{author}{\bibinfo{person}{Timo Schick} {and} \bibinfo{person}{Hinrich
  Sch{\"u}tze}.} \bibinfo{year}{2021}\natexlab{}.
\newblock \showarticletitle{Exploiting Cloze-Questions for Few-Shot Text
  Classification and Natural Language Inference}. In
  \bibinfo{booktitle}{\emph{Proceedings of EACL}}. \bibinfo{pages}{255--269}.
\newblock


\bibitem[Shin et~al\mbox{.}(2020)]%
        {DBLP:conf/emnlp/ShinRLWS20}
\bibfield{author}{\bibinfo{person}{Taylor Shin}, \bibinfo{person}{Yasaman
  Razeghi}, \bibinfo{person}{Robert~L. Logan~IV}, \bibinfo{person}{Eric
  Wallace}, {and} \bibinfo{person}{Sameer Singh}.}
  \bibinfo{year}{2020}\natexlab{}.
\newblock \showarticletitle{{A}uto{P}rompt: {E}liciting {K}nowledge from
  {L}anguage {M}odels with {A}utomatically {G}enerated {P}rompts}. In
  \bibinfo{booktitle}{\emph{Proceedings of EMNLP}}.
  \bibinfo{pages}{4222--4235}.
\newblock


\bibitem[Sun et~al\mbox{.}(2021)]%
        {DBLP:conf/acl/SunQLXZBLB20}
\bibfield{author}{\bibinfo{person}{Si Sun}, \bibinfo{person}{Yingzhuo Qian},
  \bibinfo{person}{Zhenghao Liu}, \bibinfo{person}{Chenyan Xiong},
  \bibinfo{person}{Kaitao Zhang}, \bibinfo{person}{Jie Bao},
  \bibinfo{person}{Zhiyuan Liu}, {and} \bibinfo{person}{Paul Bennett}.}
  \bibinfo{year}{2021}\natexlab{}.
\newblock \showarticletitle{Few-Shot Text Ranking with Meta Adapted Synthetic
  Weak Supervision}. In \bibinfo{booktitle}{\emph{Proceedings of ACL}}.
  \bibinfo{pages}{5030--5043}.
\newblock


\bibitem[Talmor and Berant(2019)]%
        {talmor2019multiqa}
\bibfield{author}{\bibinfo{person}{Alon Talmor} {and} \bibinfo{person}{Jonathan
  Berant}.} \bibinfo{year}{2019}\natexlab{}.
\newblock \showarticletitle{MultiQA: An Empirical Investigation of
  Generalization and Transfer in Reading Comprehension}. In
  \bibinfo{booktitle}{\emph{Proceedings of ACL}}. \bibinfo{pages}{4911--4921}.
\newblock


\bibitem[Vu et~al\mbox{.}(2020)]%
        {vu2020exploring}
\bibfield{author}{\bibinfo{person}{Tu Vu}, \bibinfo{person}{Tong Wang},
  \bibinfo{person}{Tsendsuren Munkhdalai}, \bibinfo{person}{Alessandro
  Sordoni}, \bibinfo{person}{Adam Trischler}, \bibinfo{person}{Andrew
  Mattarella-Micke}, \bibinfo{person}{Subhransu Maji}, {and}
  \bibinfo{person}{Mohit Iyyer}.} \bibinfo{year}{2020}\natexlab{}.
\newblock \showarticletitle{Exploring and Predicting Transferability across NLP
  Tasks}. In \bibinfo{booktitle}{\emph{Proceedings of EMNLP}}.
  \bibinfo{pages}{7882--7926}.
\newblock


\bibitem[Wang et~al\mbox{.}(2019)]%
        {wang2019can}
\bibfield{author}{\bibinfo{person}{Alex Wang}, \bibinfo{person}{Jan Hula},
  \bibinfo{person}{Patrick Xia}, \bibinfo{person}{Raghavendra Pappagari},
  \bibinfo{person}{R~Thomas McCoy}, \bibinfo{person}{Roma Patel},
  \bibinfo{person}{Najoung Kim}, \bibinfo{person}{Ian Tenney},
  \bibinfo{person}{Yinghui Huang}, \bibinfo{person}{Katherin Yu},
  {et~al\mbox{.}}} \bibinfo{year}{2019}\natexlab{}.
\newblock \showarticletitle{Can You Tell Me How to Get Past Sesame Street?
  Sentence-Level Pretraining Beyond Language Modeling}. In
  \bibinfo{booktitle}{\emph{Proceedings of ACL}}. \bibinfo{pages}{4465--4476}.
\newblock


\bibitem[Williams et~al\mbox{.}(2018)]%
        {williams2017broad}
\bibfield{author}{\bibinfo{person}{Adina Williams}, \bibinfo{person}{Nikita
  Nangia}, {and} \bibinfo{person}{Samuel Bowman}.}
  \bibinfo{year}{2018}\natexlab{}.
\newblock \showarticletitle{A Broad-Coverage Challenge Corpus for Sentence
  Understanding through Inference}. In \bibinfo{booktitle}{\emph{Proceedings of
  NAACL-HLT}}. \bibinfo{pages}{1112--1122}.
\newblock


\bibitem[Xiong et~al\mbox{.}(2017)]%
        {xiong2017end}
\bibfield{author}{\bibinfo{person}{Chenyan Xiong}, \bibinfo{person}{Zhuyun
  Dai}, \bibinfo{person}{Jamie Callan}, \bibinfo{person}{Zhiyuan Liu}, {and}
  \bibinfo{person}{Russell Power}.} \bibinfo{year}{2017}\natexlab{}.
\newblock \showarticletitle{End-to-End Neural Ad-hoc Ranking with Kernel
  Pooling}. In \bibinfo{booktitle}{\emph{Proceedings of SIGIR}}.
  \bibinfo{pages}{55--64}.
\newblock


\bibitem[Yang et~al\mbox{.}(2017)]%
        {yang2017anserini}
\bibfield{author}{\bibinfo{person}{Peilin Yang}, \bibinfo{person}{Hui Fang},
  {and} \bibinfo{person}{Jimmy Lin}.} \bibinfo{year}{2017}\natexlab{}.
\newblock \showarticletitle{Anserini: Enabling the use of Lucene for
  information retrieval research}. In \bibinfo{booktitle}{\emph{Proceedings of
  SIGIR}}. \bibinfo{pages}{1253--1256}.
\newblock
\urldef\tempurl%
\url{https://doi.org/10.1145/3077136.3080721}
\showURL{%
\tempurl}


\bibitem[Zhang et~al\mbox{.}(2020a)]%
        {zhang2020selective}
\bibfield{author}{\bibinfo{person}{Kaitao Zhang}, \bibinfo{person}{Chenyan
  Xiong}, \bibinfo{person}{Zhenghao Liu}, {and} \bibinfo{person}{Zhiyuan Liu}.}
  \bibinfo{year}{2020}\natexlab{a}.
\newblock \showarticletitle{Selective weak supervision for neural information
  retrieval}. In \bibinfo{booktitle}{\emph{Proceedings of The Web Conference
  2020}}. \bibinfo{pages}{474--485}.
\newblock


\bibitem[Zhang et~al\mbox{.}(2020b)]%
        {zhang2020little}
\bibfield{author}{\bibinfo{person}{Xinyu Zhang}, \bibinfo{person}{Andrew
  Yates}, {and} \bibinfo{person}{Jimmy Lin}.} \bibinfo{year}{2020}\natexlab{b}.
\newblock \showarticletitle{A little bit is worse than none: Ranking with
  limited training data}. In \bibinfo{booktitle}{\emph{Proceedings of
  SustaiNLP: Workshop on Simple and Efficient Natural Language Processing}}.
  \bibinfo{pages}{107--112}.
\newblock


\end{thebibliography}
\clearpage
\appendix
 \section{Appendices}
%\subsection{Data Partitioning}\label{appendix:data_partition}
%\input{Tables/subset_details}
%The data partitioning details of MS MARCO dataset is shown in Table~\ref{tab:data partition details}
\subsection{Prompt Selection}\label{appendix:prompt}
In this subsection, we describe how we design and generate prompts in our experiments.
Given query $q$ and document $d$, a prompt, either discrete or continuous, defines the input pattern for pre-trained language models.

\textbf{Discrete Prompts.} 
For monoT5 and \ours{}, we use manually designed prompts as described in Section~\ref{sec:method}.

For encoder-only models (BERT, RoBERTa), we use both manual discrete prompts and automatic discrete prompts, where a $(q,d)$ pair is packed by a template with a [MASK] token into a fill-in-the-blank problem.
For manual prompts, we design the template as:
\begin{equation}
\label{eq:template}
\small
    T(q,d)= \texttt{[q]}~\text{is}~[\text{MASK}]~(\text{relevant}|\text{irrelevant})~\text{to}~\texttt{[d]}~,
\end{equation}
where ``relevant'' ``irrelevant'' are label words representing $y=1$ and $y=0$, respectively. 
The prediction is based on the softmax on the [MASK] token's logits over the label words:
\begin{align}
\small
    P(y|(q,d))&=P(t=\mathcal{M}(y))=\frac{\exp (\mathbf{w}_{\mathcal{M}(y)}^{\mathsf{T}} \mathbf{h}_{t})}{\sum_{y^{\prime} \in \mathcal{Y}} \exp (\mathbf{w}_{\mathcal{M}(y^{\prime})}^{\mathsf{T}} \mathbf{h}_{t})},
\end{align}
where $t$ is the output token located in the [MASK] position, $\mathbf{h}_{t}$ is its hidden representation, and $\mathcal{M}(y)$ is the verbalizer mapping task labels into corresponding label words. 
Note that we explicitly specify the label words in the template (Eq.~\ref{eq:template}) since we find in pilot study that it works better.

We also follow \citet{DBLP:conf/acl/GaoFC20} to automatically generate discrete prompts for BERT and RoBERTa. 
We follow the ``Auto T'' setting, i.e. fix label words, and use T5 to generate templates. 
We sample 5 few-shot training datasets (2-way-16-shot) from the official training set.
Then we use T5-base to generate templates for each few-shot dataset. 
We manually specify ``relevant'' and ``irrelevant'' as our label words, then use beam search (beam width=50) to decode multiple template candidates.
Once the template candidates are generated, we perform prompt-based fine-tuning using large models (BERT-large, RoBERTa-large) on every few-shot dataset and rank the templates according to the model's best performance on the official development set. 
Then we choose the template with the highest score as our template. 
The template we choose for BERT is:
\begin{equation}
\small
    T(q,d)=~\texttt{[q]}?~\text{Is}~\text{this}~[\text{MASK}]~\text{to~your~situation~?}~\texttt{[d]}~,
\end{equation}
and for RoBERTa is:
\begin{equation}
\small
    T(q,d)=~\texttt{[q]}?~\text{Which}~\text{is}~[\text{MASK}]?~\texttt{[d]}~.
\end{equation}
% Besides, template we choose for BERT is:

\textbf{Continuous Prompts.} We follow \citet{lester2021power} to conduct prompt tuning on T5 models, which only tunes several tokens' embeddings while keeping the parameters of the PLM frozen.
%Given the probable suboptimal of discrete prompts and the lower parameter-efficiency of model-tuning, we also conduct some prompt tuning experiments using continuous prompts,which follows \citeauthor{lester2021power}'s work. 
%For our models, we keep the PLM parameters frozen, and only tune the embeddings of prompt tokens which not in .
The template that we use is like below: 
\begin{equation}
\small
    T(q,d)=\text{[$s_1$]}~\texttt{[q]}~\text{[$s_2$]}~\texttt{[d]}~\text{[$s_3$]}~,
\end{equation}
where $\text{[$s_1$]}$, $\text{[$s_2$]}$, $\text{[$s_3$]}$ are sequences consisting of tunable continuous tokens.
% , whose embeddings will be tuned during the prompt tuning process. 
In our experiments, $\text{[$s_1$]}$, $\text{[$s_2$]}$, and $\text{[$s_3$]}$ are initialized as ``Task: Find the relevance between Query and Document. Query:'' ``Document:'', and ``Relevant:'', respectively.

%Under all settings, we use the CrossEntropy Loss to tune our model.
%\subsection{Automatically generated prompts}\label{appendix:auto prompts}
\subsection{Data Partition}\label{appendix:data_partition}
In this subsection, we describe how we partition the data to simulate various training scenarios.

\textbf{MS MARCO.} 
For MS MARCO, we partition the data according to the number of training queries.
We construct four training scenarios from the official training set, containing \{5,50,1k,530k\} training queries each.
Each query in training set is paired with one relevant passage and one irrelevant passage sampled from top 1000 passages returned by BM25.
In every scenario, we sample additional queries from the official training set to set up a development set for model selection.
For scenarios in which the number of training queries is no more than 50, queries in the development set keeps a same size as the training set.
For scenarios with more than 50 training queries, the number of queries in the development set is fixed at 500 to accelerate validation process. 
Each query in the development sets is paired with top 1000 BM25-retrieved passages for reranking.

\textbf{Robust04.}
We split Robust04 into five folds, each containing 20\% of the queries.
In each fold, we sample \{0.2\%, 2\%, 100\%\} relevance labels.
Following~\citet{zhang2020little}, to construct an $r$-sampled split under total $N$ queries and $M$ associated labels, we literately drop a query until there exist less than $rM$ labels. 
Then we re-insert the last dropped query, and start to randomly remove a label per query until the number of labels reaches $rM$.

\end{document}